\documentclass{svjour2}                    % onecolumn
\smartqed  % flush right qed marks, e.g. at end of proof
\usepackage{graphicx}
\usepackage[normalem]{ulem}
\usepackage{color}

\begin{document}

\title{Interpreting superfluid spin up through the response of the container%\thanks{Grants or other notes
}


\author{C. A. van Eysden          \and
        A. Melatos %etc.
}


\institute{C. A. van Eysden \at
              School of Physics, University of Melbourne, Parkville, VIC, 3010, Australia \\
              \email{ave@unimelb.edu.au}           %  \\
           \and
           A. Melatos \at
              School of Physics, University of Melbourne, Parkville, VIC, 3010, Australia \\
              \email{amelatos@unimelb.edu.au}
}

\date{Received: date / Accepted: date}

\maketitle

\begin{abstract}

A recipe is presented for interpreting non-invasively the transport processes at work during relaxation of a cylindrical, superfluid-filled vessel, after it is accelerated impulsively and then allowed to respond to the viscous torque exerted by the contained fluid.
The recipe exploits a recently published analytic solution for Ekman pumping in a two-component superfluid, which treats the back-reaction self-consistently in arbitrary geometry for the first time.
The applicability of the recipe to He II, $\rm{^3 He}$, $\rm{^3 He}$--$\rm{^4 He}$ mixtures and Bose-Einstein condensates is assessed, and the effects of turbulence discussed.
% The experiment complements existing measurement techniques involving tuning forks, vibrating wires, and Kapton diaphragms.

\end{abstract}

\section{Introduction}
\label{epsec1}

The study of transport processes in superfluids has a long history \cite{and66,lie84,don05}.
Measurements of transport coefficients, like the viscosity or mutual friction, help to shed light on microscopic physics like the phonon-vortex and phonon-roton interactions in He II \cite{zad09}.
Transport processes are also useful for characterizing superfluids with complicated phase diagrams like $\rm{^3 He}$ \cite{leg75} and $\rm{^3 He}$--$\rm{^4 He}$ mixtures \cite{and75}, changing more strongly between phases than some thermodynamic variables like density.
In non-laminar flows, the effective viscosity offers a practical way to quantify complex dynamics like vortex tangles and turbulence \cite{sta00,cha07,wal07}.

Spin-up experiments have played a crucial role historically in the understanding of superfluidity and represent one important arena where transport processes play out \cite{wal58,hal60,pel60,rep60,cam82}.
A diverse range of phenomena were observed and explored by Tsakadze and Tsakadze \cite{tsa80}, including spin down of smooth- and rough-walled containers, quasi-periodic oscillations, and spasmodic jumps in angular velocity.
The data have never been reproduced quantitatively by a first-principles theoretical analysis; in a recent publication \cite{van11c}, we demonstrated agreement at the $0.5\%$ level between the data and a hydrodynamic theory for the smooth spin down, but the other phenomena remain unexplained
\footnote{Adams et al. \cite{ada85} explained the spin up from rest in rough-walled containers below $1.3\,{\rm K}$ by modelling pinning as friction between the vortices and the walls.}.
Subsequent experiments in helium II and $^3$He have focused on quantum turbulence \cite{bar95,don03,tsu09,vin10}.
Below a critical temperature, the energy flux in the turbulent (Kolmogorov) cascade transfers to individual vortex lines in a Kelvin-wave cascade when it reaches the quantum scale, and the viscosity drops dramatically \cite{wal07,elt07,fis09}.
Mutual friction and viscosity also play an important role in the onset of turbulence in $^3$He-$^4$He mixtures, which have been studied less thoroughly to date \cite{cas85,gri10}.

In a spin-up experiment, the container responds to the viscous torque exerted by Ekman pumping \cite{gre63}.
The latter process is sensitive to transport physics that cannot be probed in traditional, small-amplitude, second-sound or torsional-oscillator measurements of the mutual friction and viscosity.
For example, viscosity measurements using fine-wire vibration are difficult to conduct in $^3$He, where the oscillations are strongly damped.
Ekman pumping can also be used to understand turbulent flow, where the spin-up times are typically faster than laminar flow \cite{elt10b,elt07,wal11b}.
In astrophysical applications like neutron stars, remote observations are the {\it only} possible way to measure the superfluid coefficients in bulk nuclear matter \cite{alp84,van10}.
Spin-up studies are independent of experiments involving vibrational modes.

The paper is structured as follows.
In \S\ref{epsec2}, we outline the set-up of a generic spin-up experiment and the role potentially played by turbulence therein.
In \S\ref{epsec3}, an analytic solution for superfluid Ekman pumping derived by van Eysden and Melatos \cite{van11a,van11b} is used to demonstrate how to interpret the measured container response in terms of elements of the flow physics and extract accurate values for the transport coefficients.
In \S\ref{epsec4}, we discuss how Ekman pumping experiments can be conducted and interpreted in He II, $^3$He, $^3$He-$^4$He mixtures and Bose-Einstein condensates, with an emphasis on identifying the optimal regimes in which to probe various transport phenomena.

\section{Spin-up experiment}
\label{epsec2}

Consider a rigid, cylindrical vessel filled with superfluid.
To begin with, the vessel is spun up to a constant angular velocity $\Omega_s$ about the $z$-axis (unit vector ${\bf k}$), until corotation between all components is established.
The angular velocity of the vessel is then increased impulsively by a small amount $\delta \Omega $.
The subsequent motion of the vessel is tracked as accurately as possible, as it responds to the viscous torque exerted by the superfluid and the friction in the mount.
The angular velocity versus time is fitted to the spin-down curve predicted by theory \cite{van11a,van11b}.
Two exponential time-scales are identified, from which the Hall-Vinen mutual friction parameters $B$, $B'$ and shear viscosity $\eta$ can be extracted.

To analyze and interpret the post-jump response, we exploit a recently derived theoretical solution for the spin up of a superfluid-filled container \cite{van11a,van11b} based on the Hall-Vinen-Bekharevich-Khalatnikov (HVBK) equations, which model the superfluid in terms of viscous and inviscid components coupled by a mutual friction force.
The analysis assumes that the flow is linear, with Rossby number $\varepsilon=\delta\Omega/\Omega_s\ll1$.
Linearized and in the rotating frame, the HVBK equations take the dimensionless form \cite{per08,van11a}
\begin{eqnarray}
E^{1/2}\frac{\partial {\bf v}_n}{\partial \tau}+ 2 {\bf k} \times {\bf v}_n&=&-\nabla p_n +E \nabla^2 {\bf v}_n+\frac{\rho_s}{\rho} {\bf F}\, , \label{epeq1} \\
E^{1/2}\frac{\partial {\bf v}_s}{\partial \tau}+ 2 {\bf k} \times {\bf v}_s&=&-\nabla p_s-\frac{\rho_n}{\rho} {\bf F}\, , \label{epeq2} \\
\nabla\cdot{\bf v}_n&=&0\, , \label{epeq3} \\
\nabla\cdot{\bf v}_s&=&0\, . \label{epeq4}
\end{eqnarray}
where ${\bf v}_{n,s}$ and $p_{n,s}$ are the bulk velocities and pressures of the viscous $(n)$ and inviscid $(s)$ components.
In the theory, it is assumed that the container rotates with Ekman number $E\ll1$.
The Ekman number is defined as
\begin{equation}
 E=\frac{\nu_n}{ h^2 \Omega_s} \label{ep201}
\end{equation}
where $\nu_n=\eta/\rho_n$ is the kinematic viscosity of the viscous component.
The linearized mutual friction force takes the Hall-Vinen form \cite{hal56b}
\begin{equation} \label{epeq5}
 {\bf F}=B  {\bf k}\times\left[{\bf k}\times\left({\bf v}_n-{\bf v}_s\right)\right] +B' {\bf k}\times\left({\bf v}_n-{\bf v}_s\right) \,, \label{epeq5a}
\end{equation}
and $B$, $B'$ symbolize the mutual friction coefficients.
The velocity and pressure scales are chosen to be $h\delta\Omega_s$ and $\rho\Omega_s h^2 \delta\Omega_s$ respectively, where $\rho$ is the total mass density of the fluid, and $h$ is the half-height of the cylinder.

In (\ref{epeq1})--(\ref{epeq5a}), both fluid components are assumed to be incompressible and the thermal conductivity vanishes.
Therefore the only dissipative coefficient appearing in (\ref{epeq1})--(\ref{epeq4}) is the first viscosity $\nu_n$ \cite{lan59}.
The circulation in the inviscid component arises from a rectilinear array of quantized vortices polarized along the rotation axis, whose critical angular velocity for formation is \cite{vin61}
\begin{equation}
 \Omega_c=\frac{\Gamma}{R^2}\ln \left(\frac{r}{a_0}\right)\,, \label{ep401}
\end{equation}
where $r$ is the vessel radius, $a_0$ is the vortex core radius, and $\Gamma=2\pi \hbar / m$ is the quantum of circulation (Planck's constant divided by the mass of one superfluid boson).
The HVBK equations are derived by averaging smoothly over a high density of vortex lines and require the vortex line separation to satisfy $\left(\Gamma/2 \Omega_s\right)^{1/2}\ll r$ \cite{hil77,bay83,rei93,per08}.
The inviscid component spins up when vortices created at the side-walls of the cylinder migrate inwards \cite{elt10a}.
The new vortices added during the impulsive spin up are a small fraction of the total number ($\propto \varepsilon$), so the analysis is valid even in the presence of a vortex formation barrier \cite{vin10}.
In other words, even if new vortices cannot be added during spin-up and thus a vortex-free region forms close to the cylinder wall, the overall dynamics would not be significantly affected, since the vortex-free region is small.

To keep the problem analytically tractable, the walls of the container are assumed to be smooth and pinning is neglected, as contrived in many laboratory experiments, e.g., Refs. \cite{tsa80,ada85,elt10a}.
The vortex tension, parameterized by $\nu_s$, is also neglected \footnote{The tension parameter $\nu_s$ looks like a kinematic viscosity, but it is non-dissipative and gives rise to Kelvin waves.  It takes the form $\nu_s=\left(\Gamma/4\pi\right) {\rm ln}(b_0/a_0)$, where $b_0=(\Gamma/2\Omega_s)^{1/2}$ is the inter-vortex spacing.}.
We do this, not because the tension is necessarily small, but rather because $\nu_s$ {\it only} appears in the {\it inviscid} boundary layer solution \cite{rei93}; the interior flow solution and spin-up time of Ref. \cite{rei93} are recovered despite assuming $\nu_s=0$, suggesting that $\nu_s$ plays a minimal role in the spin-up of smooth-walled containers.
(See section 4.4 of Ref. \cite{van11a} and section 2.4 in Ref. \cite{van11c} for details.)
Indeed, the $\nu_s=0$ theory agrees well with experimental data for He II in smooth-walled containers in the temperature range $1.4\,{\rm K}<T<1.8\,{\rm K}$ \cite{van11c}.
In the present analysis, the viscous component obeys the usual no-slip and no-penetration boundary conditions with respect to the tangential and normal velocity respectively.
For the inviscid component, no penetration is sufficient.
The theory is not valid in rough-walled containers, where the spin up of superfluids proceeds differently \cite{tsa80,cam82,ada85,rei93}.

\begin{figure} [h!]
  \includegraphics[width=0.45\textwidth]{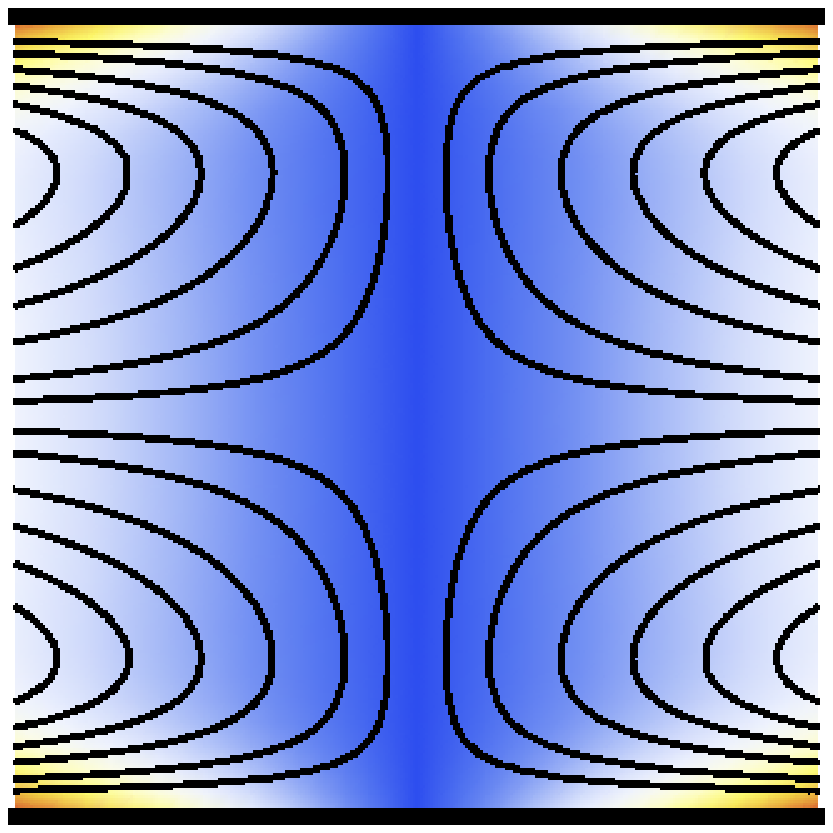}
  \includegraphics[width=0.45\textwidth]{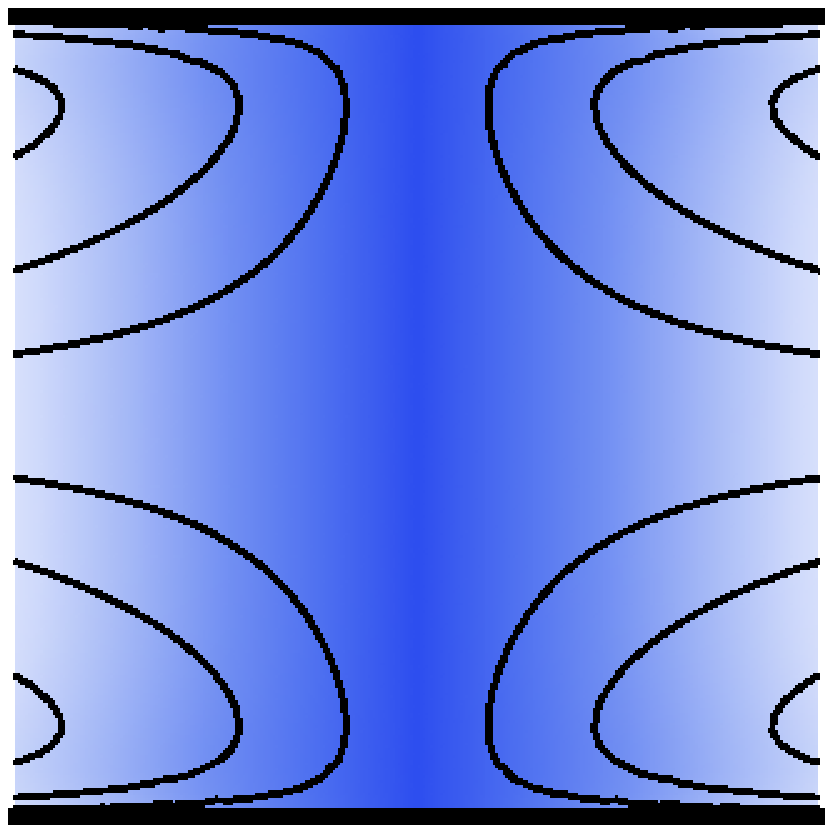}
\caption{Ekman pumping for a two-component superfluid in cylindrical geometry at $t=E^{-1/2}\Omega_s^{-1}$ for the viscous component (left panel) and the inviscid component (right panel).
Poloidal streamlines are plotted as thick black curves.
The flow is clockwise in the top-left corner and mirrored about the vertical and horizontal bisectors in the other quadrants.
The magnitude of the azimuthal velocity is represented by the shading, where red is highest and blue is lowest.
Model parameters are $B=0.8$, $B'=0.2$, $\rho_n=0.5 \rho$, $E=0.01$, $K=0.25$ and $T_0=-0.0063$.}
\label{epfig1}
\end{figure}

Angular momentum is transferred from the container to the fluid by Ekman pumping, which is illustrated in Figure \ref{epfig1}.
The flow pattern of the viscous (left panel) and inviscid (right panel) components is shown at $t=E^{-1/2}\Omega_s^{-1}$.
In the viscous component, the Coriolis force drives a secondary flow in boundary layers of thickness $E^{1/2} h$ along the top and bottom of the cylinder.
The boundary flow draws in fluid from the interior, spins it up, and cycles it back into the interior on the Ekman time-scale $E^{-1/2}\Omega^{-1}_s$.
The inviscid component is spun up by the viscous component via the mutual friction force.
When the mutual friction is strong, i.e. $B\gg E^{1/2}$, as in Figure \ref{epfig1}, the inviscid component undergoes poloidal circulation as it is dragged along by the viscous component.
For intermediate mutual friction coupling, i.e., $B\sim E^{1/2}$, the inviscid component does not circulate poloidally, and the azimuthal viscous component spins up its inviscid counterpart over the mutual friction time-scale.
For weak coupling, i.e., $B \ll E^{1/2}$, the viscous component spins up uninhibited by the inviscid component, which is spun up by mutual friction over a much longer time scale.
A detailed study of the role of mutual friction in the spin-up of two-component superfluids can be found elsewhere \cite{van11a}.

The Ekman and Rossby numbers control a number of important aspects of the experiment.
Both should be made as small as is practical, although for $E<\varepsilon^2$ turbulence may switch on, with important consequences (see below).
The scale of the apparatus $h$ enters through the Ekman number; a larger container reduces $E$.
For a container with $h=2\,{\rm cm}$ rotating at $\Omega_s=1\,{\rm rad\,s^{-1}}$ (as in Ref. \cite{tsa80}), we obtain $E \ll 1$ for most known terrestrial superfluids (or indeed fluids), including helium II.
Before the impulsive spin up, a constant rotation rate should be maintained for a few minutes (corresponding to several Ekman times; the diffusion time is hours), to ensure that the interior fluid attains steady-state co-rotation and all inertial oscillations are damped.
The Ekman number also controls the maximum duration of the spin-up event, which one contrives to be shorter than the Ekman time-scale $E^{-1/2}\Omega_s^{-1}$ \cite{van11c}.
The Rossby number must be large enough for $\varepsilon\Omega_s$ to be accurately measurable.

\subsection{Turbulence}

The above analysis also describes turbulent flow, when the transport coefficients are interpreted as eddy-averaged quantities.
Many qualitative aspects of laminar Ekman pumping persist when the flow is turbulent; e.g., the secondary flow pictured in Fig. \ref{epfig1} is still present \cite{gre68}.
Similar qualitative behaviour has also been observed for the transition to turbulent vortex flow in He II \cite{ada85,wal07}.

In a superfluid, turbulence is either quasi-classical, quantum mechanical, or a combination of the two.
Quasi-classical turbulence occurs on macroscopic scales through the formation of Kolmogorov-like eddies, even when there is no counterflow (i.e., ${\bf v}_n={\bf v}_s$ everywhere).
Quasi-classical turbulence is predicted to occur when
\begin{equation}
 {\rm Re}_{\alpha}=\frac{2-\rho_n B'}{\rho_n B}\,, \label{ep403}
\end{equation}
which is the ratio of the reactive and dissipative components of the mutual friction force, becomes large \cite{fin03,elt10b}.
Equation (\ref{ep403}) suggests that this occurs when $\rho_n\rightarrow0$ as $T\rightarrow0$.
Turbulence can also enter classical Ekman pumping, when the Ekman layer becomes unstable \cite{gre68,rei93,van11c}.
In a Navier-Stokes fluid, this occurs for
\begin{equation}
 E < \varepsilon^2\left(56.3+58.4 \varepsilon\right)^{-2}\,, \label{ep402}
\end{equation}
suggesting that $\varepsilon<E^{1/2}\ll1$ is required for laminar Ekman pumping.

Quantum turbulence arises when the counter-flow velocity along the vortex lines, $({\bf v}_n-{\bf v}_s)\cdot \hat{\bf \Omega}$, exceeds the critical value $\vert{\bf v}_{DG}\vert=2\left(2 \Omega_s \nu_s\right)^{1/2}$ \cite{gla74,don05}, at which point Kelvin waves are excited and a fully developed, self-sustaining, reconnecting vortex tangle develops \cite{tsu04,tsu09,vin10,pao11}.
This type of turbulence occurs on microscopic scales (that of the vortex lines), even if ${\bf v}_n$ and ${\bf v}_s$ are uniform on macroscopic scales.
The mutual friction force per unit mass (written in terms of starred, dimensional quantities) takes the Gorter-Mellink form \cite{gor49,per08}
\begin{equation}
 {\bf F^*}=A \frac{\rho_n\rho_s}{\rho} \left({\bf v}^*_s-{\bf v}^*_n\right)^3\,, \label{ep402aa}
\end{equation}
where $A$ is a phenomenological constant which depends on the type of fluid and temperature.
In Ekman pumping, the counter-flow velocity scales as $\varepsilon E^{1/2}\Omega_s h$ \cite{van11a}, so that we require
\begin{equation}
 \varepsilon<2\left(2 \nu_s/\nu_n\right)^{1/2}\, \label{ep402a}
\end{equation}
for laminar flow.
We check this condition in several applications in \S\ref{epsec4}.

To characterize the type of turbulence, the superfluid Reynolds number is defined \cite{hos11}
\begin{equation}
 {\rm Re}_{s}=\frac{\Omega R^2}{\nu_s} \,. \label{ep403a}
\end{equation}
Turbulence is observed in He II and $^3$He-B when both ${\rm Re}_{\alpha}>1$ and ${\rm Re}_{s}\gg1$ hold \cite{hos11}.
Quasi-classical Kolmogorov turbulence is found in the regime ${\rm Re}_{\alpha}<{\rm Re}_{s}^{1/2}$, while quantum turbulence is found when ${\rm Re}_{\alpha}>{\rm Re}_{s}^{1/2}$.
The turbulent state depends on the particular fluid in question; we discuss some particular examples in \S\ref{epsec4}.

An important issue is how the dissipation changes between laminar and turbulent flow and how this carries through to the observed spin-down time of the container.
Typically, turbulence enhances disspiation.
This is observed in spin-down and spin-up experiments in $^3$He-B, where the turbulent response is faster at a given temperature \cite{elt07,elt10b,wal11b}.
Interestingly, experiments in He II show that the spin-down time is longer than expected for laminar flow near the superfluid transition temperature, where turbulence is predicted \cite{van11c}.
Dissipation can also change significantly within the turbulent state.
Below a critical temperature, the energy flux in a turbulent (Kolmogorov) cascade transfers to individual vortex lines in a Kelvin-wave cascade when it reaches the quantum scale, and the viscosity drops dramatically \cite{wal07,elt07,fis09}.

\section{Container response} \label{epsec3}

\subsection{Spin-down curve}
\label{epsec3a}

Let $\Omega(\tau)=\Omega_s+\delta\Omega f(\tau)$ denote the angular velocity of the vessel at $\tau\geq0$, where the dimensionless variable $\tau$ is the usual time-coordinate normalized by the Ekman time $E^{-1/2}\Omega_s^{-1}$ \cite{gre63,van11a}.
At time $\tau=0$, when the cylinder is released after being accelerated impulsively, the initial velocities of the viscous and inviscid components are zero in the rotating frame, and $f(0)$ equals unity.
The constant frictional torque in the apparatus is denoted $T_0$ in dimensionless units and is negative.
For a rigid, symmetric container, the integral equation of motion is \cite{van11b}
\begin{equation}
 f\left(\tau\right)=-K \int_0^\tau {\rm d\tau'} \, \left[ \dot{g}^A(\tau-\tau')+\dot{g}^B(\tau-\tau')\right] f(\tau')+ T_0 \tau  + 1 \,. \label{ep301}
\end{equation}
Equation (\ref{ep301}) captures self-consistently the back-reaction of the superfluid on the container and vice versa, e.g. by including the continuously changing boundary conditions on the superfluid, as $\Omega(\tau)$ evolves.
For a cylinder of radius $r$, height $2 h$, and moment of inertia $I_c$, the ratio of the nominal moment of inertia of the fluid (rotating as if it were a rigid body of uniform density $\rho$) to the moment of inertia of the container is given by
\begin{equation}
 K=\frac{\rho \pi r^4 h}{I_c}\,.\label{ep301a}
\end{equation}
In (\ref{ep301}), we make the definitions
\begin{eqnarray}
 g^A(\tau)&=& \frac{J \left(  e^{\omega_+\tau}-e^{\omega_-\tau} \right)}{\omega_+-\omega_-} \,,  \label{ep302} \\
 g^B(\tau)&=&\frac{ \omega_- \left(e^{\omega_+\tau}-1\right)-\omega_+\left( e^{\omega_-\tau}-1\right) }{\omega_+-\omega_-} \label{ep303} \,,
\end{eqnarray}
where the overdot symbolizes a derivative with respect to $\tau$, and we have
\begin{eqnarray}
 \omega_{\pm}&=&-\left(\frac{\beta+I}{2}\right)\pm \left[\left(\frac{\beta+I}{2}\right)^2-\beta J \right]^{1/2} \, ,  \label{ep304} \\
 \beta&=&\frac{2 B E^{-1/2}}{2-B'} \, , \label{ep305} \\
 I&=&\frac{2 \lambda_{+}}{\lambda_{+}^2+\lambda_{-}^2 } \,,\label{ep306} \\
 J&=&\rho_n \lambda_{-} \,,\label{ep307}
\end{eqnarray}
\begin{equation}
 \lambda_{\pm}=\left\{ \left[\frac{\left(B'-2\right)^2+B^2}{\left(\rho_n B'-2 \right)^2+\left(\rho_n B\right)^2}\right]^{1/2}  \mp \frac{2 \rho_s   B}{\left(\rho_n B'-2\right)^2+\left(\rho_n B\right)^2 }\right\}^{1/2} \,. \label{ep308}
\end{equation}

\begin{figure}[h!]
  \includegraphics[width=0.45\textwidth]{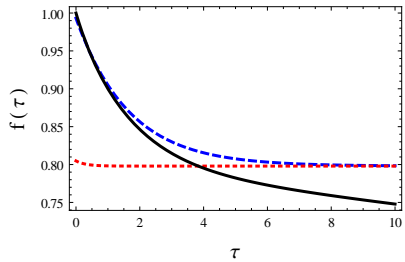}
\caption{Spin down of a superfluid-filled container accelerated impulsively at $\tau=0$.
The thick black curve corresponds to (\ref{ep309}), while the first and second terms in (\ref{ep309}) are graphed separately as blue-dashed and red-dotted curves respectively.
The model parameters are the same as in Figure \ref{epfig1}.}
\label{epfig2}
\end{figure}

Equation (\ref{ep301}) has the exact solution
\begin{equation} \label{ep309}
 f(\tau)=\left(1-C-f_\infty \right)\exp\left(\varpi_+\tau\right)+C\exp\left(\varpi_-\tau\right)+ D \tau +f_\infty \,,
\end{equation}
with the auxiliary definitions
\begin{eqnarray}
 \varpi_\pm&=&-\left(\frac{\beta+I+J K}{2}\right) \pm \left[\left(\frac{\beta+I+J K}{2}\right)^2- \beta J \left(1+K\right) \right]^{1/2}\,, \label{ep310}\\
 C&=&\frac{ K \varpi_+  \left(\varpi_- + \beta   \right)\left(\varpi_-+T_0\right)}{\left(1 + K\right) \varpi_- \beta \left(\varpi_+ - \varpi_-\right)} \,, \label{ep311} \\
 D&=&\frac{T_0}{1+ K} \,, \label{ep313} \\
 f_\infty&=&\frac{1}{1+ K}\left[1+T_0\left(\frac{\varpi_- +\varpi_+}{\varpi_- \varpi_+}-\frac{\omega_-+ \omega_+}{\omega_- \omega_+}\right)\right] \,. \label{ep312}
 \end{eqnarray}
Equation (\ref{ep309}) can be verified by substitution into (\ref{ep301}).
It comprises two exponential decays, which depend on the mutual friction, viscosity, and the relative inertia of the container; a linear spin down, in response to the constant external torque; and a steady-state solution, which the system would relax to in the absence of an external torque.
The solution (\ref{ep309}) is graphed as a thick black curve in Figure \ref{epfig2}.
The two exponential decaying terms in (\ref{ep309}) are also graphed separately as blue-dashed and red-dotted curves respectively.
For these parameters, both decays are clearly and separately evident in the overall recovery.

The detailed form of the spin down depends on the strength of the mutual friction; see Ref. \cite{van11a} for a detailed discussion.
%In general, the decay times $\Omega^{-1}E^{-1/2} \varpi_\pm^{-1}$ are a complex mixture of mutual friction, viscosity and the relative inertia of the container.
When the superfluidity coefficients satisfy $\beta \gg1$, the two fluid components are drawn together by mutual friction much faster than the Ekman time.
In this limit, we obtain $-\Omega_s^{-1} E^{-1/2} \varpi_+^{-1}= E^{-1/2} \Omega_s^{-1} J (1+K)/h$, which is the superfluid Ekman time \cite{rei93,van11a}, while $-\Omega_s^{-1}E^{-1/2} \varpi_-^{-1}=B \Omega_s^{-1}$ is the coupling time-scale for the two components due to mutual friction.
We also obtain $C \gg 1-C-f_\infty $, and the spin down is dominated by the first term in (\ref{ep309}).
This strong coupling limit is discussed further in \S\ref{epsec3c}.
For intermediate (i.e., $\beta\sim1$) coupling, both exponentials in (\ref{ep309}) are important, and $-\Omega_s^{-1}E^{-1/2} \varpi_{\pm}^{-1}$ are complex mixtures of the Ekman and mutual friction coupling times.

The spin-down time of the container is strongly dependent on $K$.
Increasing $I_c$, e.g. by attaching a metal disk to the spindle on which the container rotates, reduces $K$ and increases the spin-down time.
This may be desirable if the container response is too rapid for the shape of the spin-down curve to be measured accurately.
As an independent check, the experiment can be repeated for different values of $K$.

The solution (\ref{ep309}) assumes that the viscous and inviscid fluid components are initially co-rotating.
As a consequence, the pre-factors $1-C-f_\infty $ and $C$ are always positive, and the spin down is monotonic.
In general, either $1-C-f_\infty$ or $C$ can be negative, when the inviscid and viscous components initially rotate differentially.
This can cause an overshoot: the angular velocity of the container drops below the steady-state value before rising again.
Although this possibility is interesting, especially in astrophysical applications \cite{van10}, we do not consider it here, because the initial velocities of the HVBK components are difficult to fine-tune experimentally.

Dual exponential recovery is characteristic of the above system, irrespective of geometry.
Qualitatively similar behaviour is seen in a sphere, except that the shape of the spin-down curve is quasi-exponential (strictly, a latitude-weighted sum of exponentials) \cite{van11a}.
The hydrodynamic torque in non-parallel-plate geometries varies along the boundary, and solutions to (\ref{ep301}) cannot be written in analytic form like (\ref{ep309}).
The solution governing equations for $f(\tau)$ in general geometry is given in the Appendix.

\subsection{Extracting $B$, $B'$ and $E$ from the spin-down curve}
\label{epsec3b}

The analytic solution (\ref{ep309}) can be inverted, so that the superfluid transport coefficients $B$, $B'$ and $E$ can be inferred from the experimentally measured spin-down curve $\Omega(t)$.
From (\ref{ep309})--(\ref{ep312}), we obtain
\begin{eqnarray}
 \left(\frac{d f}{d \tau}\right)_{\tau=0}&=&-K J + T_0 \,, \label{ep314}\\
 \varpi_+ \varpi_-&=&\beta J (1+K) \,, \label{ep315} \\
 \varpi_+ + \varpi_-&=&-\left(\beta+I+J K\right)\,. \label{ep316}
\end{eqnarray}
Restoring dimensions according to $\varpi^*_\pm=\varpi_\pm \Omega_s E^{1/2}$, $t=E^{-1/2}\Omega^{-1}_s \tau$, and $T^*_0=\delta \Omega \Omega_s  E^{1/2} I_c T_0$,
equations (\ref{ep314})--(\ref{ep316}) can be rearranged to give
\begin{eqnarray}
 J E^{1/2}&=&\left(\delta\Omega \Omega_s K\right)^{-1}\left[\frac{T_0^*}{I_c}-\left( \frac{d \Omega}{d t}\right)_{t=0} \right]\,,\label{ep319} \\
 \beta E^{1/2}&=&\varpi^*_+ \varpi^*_-\left[\frac{K \delta \Omega }{\left(1+K\right)\Omega_s}\right]\left[\frac{T_0^*}{I_c}-\left( \frac{d \Omega}{d t}\right)_{t=0} \right]^{-1}\,, \label{ep320} \\
 I E^{1/2}&=&-\varpi^*_+ \varpi^*_-\left[\frac{K \delta \Omega }{\left(1+K\right)\Omega_s} \right] \left[\frac{T_0^*}{I_c}-\left( \frac{d \Omega}{d t}\right)_{t=0} \right]^{-1} \nonumber \\
 &&-\left(\delta\Omega \Omega_s \right)^{-1}\left[\frac{T_0^*}{I_c}-\left( \frac{d \Omega}{d t}\right)_{t=0} \right]-\frac{\left(\varpi^*_+ + \varpi^*_-\right)}{\Omega}  \,. \label{ep321}
\end{eqnarray}
The left-hand sides of equations (\ref{ep319})--(\ref{ep321}) feature the unknowns $B$, $B'$ and $E$, while the right-hand sides feature measurable quantities.
For example, when the container is empty, its spin-down rate is $d \Omega/d t=T_0^*/I_c$, giving the external torque in the apparatus.
When the container is full, we require the container to accelerate slowly, with $T_0^*/I_c(1+K) \Omega_s^2 E^{1/2}\ll1$ to satisfy the linearity assumption.

\begin{figure} [h!]
  \includegraphics[width=0.45\textwidth]{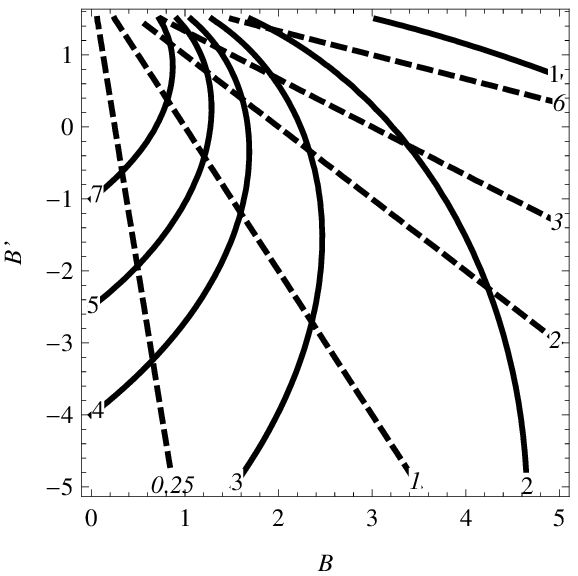}
  \includegraphics[width=0.45\textwidth]{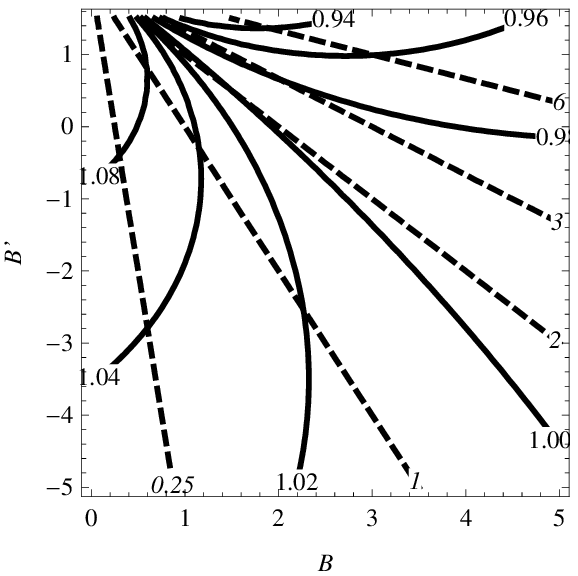}
\caption{Contours of $\beta E^{1/2}$ (dashed curves) and $I/J$ (solid curves) for $\rho_n=0.1$ (left panel) and $\rho_n=0.9$ (right panel).
The mutual friction coefficients $B$ and $B'$ can be read off easily from the intersection of the $\beta E^{1/2}$ and $I/J$ contours, measured from the spin-down curve.}
\label{epfig3}
\end{figure}

The quantities $\beta E^{1/2}$ and $I/J$ depend only on $B$, $B'$ and $\rho_n$.
If $\rho_n$ is known by some other means (e.g. published data from previous experiments, such as Ref. \cite{das57}), equations (\ref{ep320}) and (\ref{ep321}) divided by (\ref{ep319}) are sufficient to extract $B$ and $B'$.
This is done graphically in Figure \ref{epfig3}.
Contours of constant $I/J$ are plotted as solid black curves on the $B$-$B'$ plane, while $\beta E^{1/2}$ contours are plotted as dashed curves on the same plane.
Values of $I/J$ and $\beta E^{1/2}$ are calculated from equations (\ref{ep319})--(\ref{ep321}).
Given measurements of the quantities in the right-hand sides of (\ref{ep319})--(\ref{ep321}), $B$ and $B'$ can be read off the vertical and horizontal axes at the intersection point of the curves.
Once $B$ and $B'$ have been found, $E$ is determined uniquely from (\ref{ep319}) or (\ref{ep321}).

Cylindrical geometry is used here as an example, because a concise analytic solution can be obtained.
Solutions in general geometry also exist, but they do not simplify like in (\ref{ep319}) or (\ref{ep321}), so the fit must be done by adjusting $\eta$, $B$ and $B'$ by trial and error \cite{van10}.
The solution to (\ref{ep309}) in general geometry is presented in the Appendix.

\subsection{Strong coupling}
\label{epsec3c}

Many superfluids are strongly coupled by mutual friction, in the sense that one has $\beta\gg1$.
For example, He II has $1.526\leq B \leq 12.693$ for $1.30\,{\rm K}\leq T\leq 2.1767\,{\rm K}$ giving $\beta\gg1$ in the Ekman pumping regime $E\ll 1$ \cite{don11}.
In this limit, (\ref{ep301}) reduces to
\begin{eqnarray}
 f(\tau)&=&\left(1+K\right)^{-1}\left\{K\left[1-\frac{T_0 }{J\left(1+K\right)} \right]\exp\left(\varpi_+\tau\right) \right. \nonumber \\
&& \left. + T_0 \tau +1+\frac{T_0 K}{J\left(1+K\right)}  \right\}\,. \label{ep324}
\end{eqnarray}
With one of the exponential decays now imperceptible, there are fewer measurable quantities available to constrain $B$, $B'$ and $E$ [unless
the $\exp(\varpi_- \tau)$ term can be resolved by a sufficiently sensitive experiment].
Equation (\ref{ep324}) does not contain $I$ and $\beta$; all the dependence on $B$ and $B'$ resides in $J$.
Restoring dimensional variables, there is now only one relation between $B$, $B'$ and $E$, namely (\ref{ep319}).

\section{Applications}
\label{epsec4}

\subsection{Helium II} \label{epsec4a}

Helium II is the most extensively studied superfluid.
Its transport coefficients have been measured and independently verified in several ways.
Table \ref{tab1} summarizes a representative sample, noting the experimental technique and temperature range $T$.
The entries draw on information published electronically by Donnelly \cite{don11}.
Historically, viscosity measurements have been attempted with rotating cylinders \cite{woo63,don88}, damped fine-wire vibration \cite{tou63,goo73}, torsional quartz crystals \cite{web72}, and torsional oscillators \cite{wan90}.
The rotating cylinder viscometer measures the shear viscosity $\eta$ directly, whereas the other methods measure the kinematic viscosity $\eta/\rho_n $, so that $\rho_n$ must be independently measured to extract $\eta$.
The methods agree within experimental errors.
With regard to mutual friction, the $B$ coefficient is determined from the excess attenuation of second sound due to the presence of vortices in a rotating container \cite{hal56a,ben67,luc70,mat76,mil78}, while $B'$ is measured by the removal of a resonant degeneracy in a rotating second-sound cavity \cite{sny66,luc70,mat76}; see Ref. \cite{bar83} for a review.
Values for $B$ and $B'$ were compiled by Barenghi et al. \cite{bar83} and are also available electronically in Ref. \cite{don11}.

The Ekman technique, described in \S\ref{epsec3b}, for extracting $\eta$, $B$, $B'$ from spin-down curves cannot compete with the techniques in Table \ref{tab1} for helium II, as it is less accurate, and spin-up experiments are more finicky.
However it does possess two virtues.
First, it is physically independent of the other techniques, relying on Ekman pumping instead of vibrational modes, and can therefore serve as a cross-check in new situations where the data are still uncertain.
Second, it is sensitive to nonlinear phenomena like turbulence, which has been studied recently in experiments \cite{bar95,don03,tsu09,vin10} and is usually (intentionally) not excited in experiments involving small-amplitude vibrational modes.
We enlarge on this second point below.

\begin{table}
\caption{Measurements of transport coefficients in helium II}
\label{tab1}     
\begin{tabular}{lccl}
\hline\noalign{\smallskip}
Method & Quantity & $T$ (K)  & Refs.  \\
\noalign{\smallskip}\hline\noalign{\smallskip}
Oscillating disks &$\eta$ & ($ 1.2, 2.0)$ & Dash \& Taylor \cite{das57} \\
Couette viscometer &$\eta$ & ($ 0.78, 2.079)$ &  Woods \& Hollis Hallett \cite{woo63} \\
Vibrating wire & $\eta/\rho_n $ &  $(1.52, 2.16)$ &  Tough et al. \cite{tou63} \\
Vibrating wire & $\eta/\rho_n $ &  $(1.2, 4.2)$ &  Goodwin \cite{goo73} \\
Quartz crystal & $\eta/\rho_n $ &  $(1.75, 2.195)$ & Webeler \& Allen \cite{web72} \\
Torsional oscillators & $\eta/\rho_n $ &  $(1.8, 4.4)$ & Wang et al. \cite{wan90} \\
Quartz tuning fork & $\eta/\rho_n $ &  $(0.1,2.2)$ & Zadorozhko \cite{zad09} \\
\noalign{\smallskip}\hline\noalign{\smallskip}
Second sound & $B$ & $(1.289, 2.060)$ & Hall \& Vinen \cite{hal56a} \\
Second sound & $B'$ & $(1.28, 2.156)$ & Snyder \& Linekin \cite{sny66} \\
Second sound & $B$ & $(1.2, 2.08)$ & Bendt \cite{ben67} \\
Second sound & $B$, $B'$ & $(1.383, 2.167)$ & Lucas \cite{luc70} \\
Second sound & $B$, $B'$ & $(1.75, 2.167)$ & Mathieu et al. \cite{mat76} \\
Second sound & $B$ & $(1.40, 2.05)$ & Miller et al. \cite{mil78} \\
\noalign{\smallskip}\hline
\end{tabular}
\end{table}

At low temperatures, where the viscous fraction is small, one has ${\rm Re}_{\alpha}\sim10^3$ and turbulence is ubiquitous in He II \cite{wal07}.
Below a critical temperature, the energy flux in Kolmogorov cascade reaches down to the quantum scale and a Kelvin-wave cascade forms, dropping the effective viscosity by an order of magnitude \cite{wal07}.
For $T<0.5\,{\rm K}$, the viscous component also becomes a gas of ballistic quasi-particles, and the HVBK description no longer applies \cite{she08}.
% Therefore the present techniques proposed for extracting the mutual friction parameters will not apply at low temperatures in He II.

As the temperature approaches the lambda point, $\nu_n$ decreases.
Depending on $\varepsilon$, $\Omega_s$ and $h$, the condition (\ref{ep402}) may be met.
For the experiments of Tsakadze and Tsakadze, where $h=2\,{\rm cm}$, $\Omega_s\sim 3\,{\rm rad\,s^{-1}}$ and $\varepsilon\sim0.15$, turbulence is expected for $T>1.6\,{\rm K}$ \cite{van11c}.
Decreasing $h$, $\Omega_s$ and $\varepsilon$ excites turbulence at higher temperatures.
In He II, we have $0.5<\nu_s/\nu_n<6.1$ for $1.3\,{\rm K}<T<2.1\,{\rm K}$, varying approximately linearly in between \cite{don11}.
Hence, the Glaberson-Donnelly instability (\ref{ep402a}) is not excited for the experiments in Ref. \cite{tsa80}, and a window exists where laminar flow is possible in He II.

Recently, excellent agreement at the $0.5\%$ level has been obtained between the theory of Refs. \cite{van11a,van11b} and the experiments of Tsakadze \cite{tsa80} in the temperature range $1.4\,{\rm K}<T<1.8\,{\rm K}$ \cite{van11c}.
Agreement is obtained using the generally accepted values of $\eta$, $B$ and $B'$ from Ref. \cite{don11}.
This is strong evidence that the flow is laminar in this temperature range, and that the HVBK equations accurately describe the flow.
%Therefore, in principle, measurements of the spin-down of superfluid-filled containers can be fitted to the theory of Refs. \cite{van11a,van11b} to extract $\eta$, $B$ and $B'$ in He II.
In this way, the spin-down analysis in \S\ref{epsec3} complements the traditional approaches in Table \ref{tab1} by indirectly probing the conditions for turbulence in such systems.

\begin{figure}[h!]
  \includegraphics[width=0.45\textwidth]{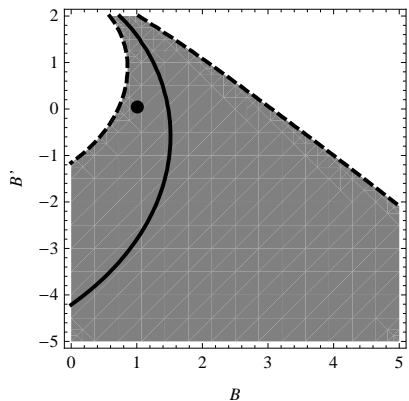}
\caption{Constraint on the mutual friction coefficients from a hypothetical measurement.  
Contours of $J E^{1/2}=0.002\pm0.0002$ for $E=9.11\times10^{-6}$ (shaded region between curves) and $J E^{1/2}=0.002$ (thick black curve).
The values of $B$ and $B'$ from the adopted database in Ref. \cite{don11} are plotted as a black dot.}
\label{epfig4}  
\end{figure}

In He II, the mutual friction is strong [$\sim O(1)$] \cite{don11}.
Therefore, only one time-scale is observed, and $B$, $B'$ and $E$ cannot be inferred uniquely.
However, if the kinematic viscosity is known from other experiments \cite{don11}, we can place useful constraints on $B$ and $B'$.
We illustrate how in Figure \ref{epfig4} with an arbitrary example.
Suppose that, from a spin-down experiment in He II at $T=2.0\,{\rm K}$, we measure $J E^{1/2}=0.002\pm0.0002$ using (\ref{ep319}).
We find $E=9.11\times10^{-6}$ using Ref. \cite{don11} (corresponding to a container with $h=2\,{\rm cm}$ rotating with $\Omega_s=5\,{\rm rad\,s^{-1}}$).
The thick black curve corresponds to $J E^{1/2}=0.002$, while the shaded region bounded by the dashed curves represents the range $J E^{1/2}=0.002\pm0.0002$ for $E=9.11\times10^{-6}$.
The generally accepted values of $B=1.008$ and $B'=0.043$ at $T=2.0\,{\rm K}$ are marked as a black dot on the figure.

\subsection{$^3$He-B} \label{epsec4b}

\begin{table}
\caption{Transport coefficient measurements in $^3$He}
\label{tab2}      
\begin{tabular}{lccccl}
\hline\noalign{\smallskip}
Method & Phase & Quantity & $T$ (mK) & $B_0$ (T)  &  Refs.  \\
\noalign{\smallskip}\hline\noalign{\smallskip}
Vibrating wire &A, B &$\eta/\rho_n$ &  $(1, 100)$ & $(0, 0.9)$ &   Alvesalo et al. \cite{alv75} \\
Torsional oscillator &A, B &$\eta/\rho_n$ &  $(1, 100)$ & $(0, 0.9)$ &  Alvesalo et al.  \cite{alv75} \\
Vibrating wire &B & $\eta/\rho_n$ & $ (0.6, 2.49) $ & 0 & Carless et al.  \cite{car83a} \\
Torsional oscillator & A, B & $\eta/\rho_n$ &  $(2.11, 2.64) $  & 0 & Main et al.  \cite{mai76} \\
First sound & B & $\eta$ & $(1.2, 2.64 )$ & 0 & Eska et al.  \cite{esk80} \\
Torsional oscillator & A & $\eta/\rho_n$ & $(2.1, 2.64 )$ & 0 & Hook \cite{hoo89a} \\
Vibrating wire & A & $\eta/\rho_n$ &  $(1.85, 2.64) $ & $(0.5, 9.2)$ & Hata et al. \cite{hat89} \\
Torsional oscillator & A & $\eta/\rho_n$ &  $(0.3, 3)$ & $(0.6,12)$ & Akimoto et al. \cite{aki95} \\
Torsional oscillator & B & $\eta/\rho_n$ &  $(0,5) $ & 0 & Nakagawa et al. \cite{nak96} \\
Torsional oscillator & A & $\eta/\rho_n$ &  $(1, 80) $ & $(0.5, 14.6)$ & Roobol et al. \cite{roo97} \\
\noalign{\smallskip}\hline\noalign{\smallskip}
Kapton diaphragm & A, B & $B$, $B'$  & $ (0.5, 2.49) $  & 0 & Bevan et al. \cite{bev97} \\
Time of flight & B & $B$ & $ (1, 2.49) $  & 0 & Finne et al. \cite{fin04} \\
\noalign{\smallskip}\hline
\end{tabular}
\end{table}

Measuring transport coefficients in superfluid $^3$He is additionally problematic because of the complicated texture physics, arising from spin interactions between $^3$He atoms \cite{vol02}.
Superfluid $^3$He can exist in an anisotropic A phase or an isotropic B phase.
%, depending on the temperature, pressure and applied magnetic field.
% In the absence of magnetic fields, the A phase forms when $\geq$21 bar of pressure is applied.
% For magnetic fields above $0.5$ T, the B phase is heavily suppressed.
Transport coefficients in the anisotropic A phase are usually formulated as tensor quantities; different tensor components are measured in different experimental situations.
% The component of the viscosity tensor measured is dependent on the technique, and the reader should refer to the appropriate reference for details.
% The techniques used to measure viscosity in helium II have also been applied to $^3$He, whose normal component is highly viscous \cite{bev97,fin04}.
Table \ref{tab2} summarizes a representative sample of viscosity and mutual friction measurements, noting the material, phase, experimental technique, temperature range $T$, and magnetic field strength $ B_0$.
The different methods agree within experimental errors \cite{roo97}.

In $^3$He, the mutual friction cannot be measured using second sound, because acoustic waves are overdamped by the high viscosity \cite{bev97,fin04}.
Alternative approaches include measuring the normal modes of transverse vibration of a Kapton diaphragm, which is used to excite a superflow \cite{hoo95},
 and measuring the time of flight of a vortex injected into a rotating cylindrical vessel \cite{fin04}.
The resonant response of quartz tuning forks in $^3$He has been investigated.
As yet, however, tuning-fork experiments have not yielded accurate values for the viscosity or mutual friction \cite{bla07,bla07b}.

In $^3$He-B, turbulence is inhibited by a large viscosity \cite{fin03,vin10}.
Noninvasive nuclear magnetic resonance measurements show that, in a cylinder, spin down is laminar all the way down to temperatures below $0.2\,T_c$, despite ${\rm Re}_{\alpha}\sim10^3$ \cite{elt10b}.
For $^3$He-B, one has $\nu_s\sim10^{-7}\,{\rm m^2\,s^{-1}}$.
Hence, from (\ref{ep402a}), we require $\varepsilon<0.02$ to prevent the onset of the Donnelly-Glaberson instability.
Also, in $^3$He, vortex cores are at least $10^2$ times larger than in He II, and surface pinning and friction do not have a substantial impact on flows in smooth-walled containers \cite{elt10a}.
However, like He II, the viscous component of $^3$He becomes ballistic in the low-temperature limit and the HVBK description breaks down.

For $^3$He-B we have approximately $5\times 10^{-4}\,{\rm m^2\,s^{-1}}<\nu_n<2.5\times 10^{-4}\,{\rm m^2\,s^{-1}}$ for $0.4<T/T_c<0.8$ \cite{nak96}, approximately $10^4$ times larger than in He II.
Therefore, we require $\Omega_s h^2\gg5\times 10^{-4}\,{\rm m^2\,s^{-1}}$ for Ekman pumping to occur ($E\ll1$).
This condition is met in typical $^3$He-B experiments.
For example in Ref. \cite{elt10a,elt10b}, one has $h\approx10\,{\rm cm}$, $\Omega_s\approx 1\,{\rm rad\,s^{-1}}$, and hence $E\sim0.02$.
Therefore $^3$He-B is amenable to the interpretative analysis described in \S\ref{epsec2} and \S\ref{epsec3}.

\subsection{$^3$He-$^4$He mixtures} \label{epsec4c}

Equations (\ref{epeq1})--(\ref{epeq5a}) describe general two-component fluid mixtures such as $^3$He and $^4$He just as well as an HVBK superfluid \cite{and06,pri04,hil77}.
To date, the study of $^3$He and $^4$He mixtures has focused on their thermodynamic properties, for applications such as cryogenic refrigeration \cite{mil01,cha10}, and the transition to turbulence \cite{gri10}.
However, transport coefficients and turbulence in $^3$He-$^4$He mixtures have been studied to some extent in the temperature range $0.15\,{\rm K}<T<2.17\,{\rm K}$, where dilute concentrations of $^3$He in the normal liquid phase are added to superfluid $^4$He.
Castelijns et al. \cite{cas85} investigated a $6.6\%$ mixture of $^3$He in $^4$He for $T<150\,{\rm mK}$ and concluded that $^3$He generates a vortex tangle in $^4$He in the same way that the normal component generates a vortex tangle in the superfluid component in He II.
The resulting mutual friction has a Gorter-Mellink form (\ref{ep402aa}), with $A\sim1.5\times10^4\,{\rm kg^{-1}\,m\,s}$, $\sim30$ times larger than a vortex tangle in pure $^4$He.
In mixtures, $^3$He also concentrates in the vortex cores of the $^4$He, which may alter the dynamics significantly.
Qin et al. \cite{qin92} measured the shear viscosity of mixtures ranging from $1$--$5\%$ $^3$He and observed that the shear viscosity decreases with increasing $^4$He concentration.
However, comparison with independent viscosity measurements using capillary flow \cite{sta60,kue72} and torsional oscillators \cite{web69,web72} yielded discrepancies of nearly $100\%$ for $1\%$ concentrations.
Spin-up experiments sensitive to the independent Ekman process may help to resolve this controversy.

Recent studies suggest that $5\%$ concentrations of $^3$He in $^4$He enhance laminar flow stability and increase the critical velocity for the transition to turbulence below $1\,{\rm K}$ \cite{gri10}.
This suggests that the spin-down experiments proposed in this paper apply to $^3$He-$^4$He mixtures.
For example, the distinctive rise in the spin-down time of the container with the onset of turbulence could serve as a useful diagnostic.

\subsection{Bose-Einstein condensates} \label{epsec4d}

Dilute-gas Bose-Einstein condensates (BECs) are an interesting application, as they are less noisily coupled to their environment than liquid helium \cite{jac06}.
Condensates rotating in magneto-optical traps form vortices which do not pin except when forced, e.g. by an optical lattice \cite{min09}; traps therefore constitute smooth-walled containers.
Another feature of the hydrodynamic analysis is that it can be applied to containers with arbitrary axisymmetric geometry, such as the parabolic traps used to construct BECs.
The general theory is summarized in the Appendix.
However, because the trap interacts with the whole BEC simultaneously and not just with its edges, no Ekman boundary layer forms.
The theory in Ref. \cite{van11a} still applies; one may imagine a situation in a two-component BEC where one component is spun up, leaving the other to respond via mutual friction.
In a BEC, the vortex core size approaches the inter-vortex spacing \cite{vin10}.
It is hard to create a rapidly rotating BEC with many vortices, because the condensate becomes unstable as the rotation frequency $\Omega_s$ approaches the trap oscillation frequency $\omega_\perp$.
Also, as $\Omega_s\rightarrow \omega_\perp$, the vortex core size grows until it saturates at the inter-vortex spacing and the vortices overlap \cite{fis03}.
Nevertheless, states with $\Omega_s\approx0.995 \omega_\perp$ containing hundreds of vortices have been constructed successfully \cite{abo01,cod03,sch04,fet10}, for which the HVBK approximation may be suitable \cite{cod03,sch04}.

\subsection{Neutron superfluid: weak coupling} \label{epsec4e}

Recently, much attention has been devoted to the thermodynamic phases and transport coefficients of bulk nuclear matter, which is generally superfluid and even superconducting \cite{alf08}.
This flowering of interest has its origin in the desire to test the theory of quantum chromodynamics in the MeV-energy, many-body regime.
The hydrodynamics of nuclear matter in beta equilibrium involves 17 transport coefficients in general \cite{and06}.
The viscosity has been measured in the high-energy (GeV), few body ($\sim$400 nucleons) regime by Au + Au collision experiments in relativistic heavy-ion colliders \cite{adl03,ada07}.
The other coefficients have not been measured.
Indeed, at present, the only practical way to measure $B$ and $B'$ is by studying the phenomenon of rotational glitches in neutron stars \cite{and05,van10}, where Nature arranges a facsimile of the spin-up experiment discussed in this paper.
Neutron star measurements of nuclear transport coefficients are used to investigate the existence of exotic nuclear superfluid phases like color-flavor locked and two-flavor superconductors \cite{mad00,gle01,man08,alf09}.

Weakly coupled two-component superfluids are believed to exist in the interiors of neutron stars,
where an ultra-dense superconducting proton-electron plasma and a neutron condensate play the role of the viscous and inviscid components respectively \cite{bay71,pin85,men91,and01,van10}.
Estimated densities range from $10^{11}\,{\rm g\,cm^{-3}}$ near the surface to $10^{14}\,{\rm g\,cm^{-3}}$ or higher in the core.
Theoretically, it is thought that the viscosity in the star arises from either neutron-neutron or electron-electron scattering, which give $E\approx10^{-7}$ \cite{flo79,cut87}, while the mutual friction force arises from inelastic electron scattering off neutron vortex cores, typically giving $10^{-5} < B < 10^{-4}$ and $B'\approx B^2$ \cite{alp84,men91,van10}.

\begin{figure}[h!]
  \includegraphics[width=0.45\textwidth]{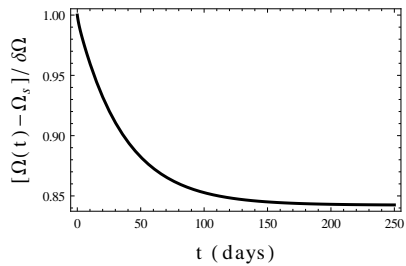}
\caption{Observed spin-down curve of the Vela neutron star after the 1985 glitch. The error bars are comparable to the width of the line; see \S\ref{epsec4e}.}
\label{epfig5}
\end{figure}

The recovery of a neutron star after an impulsive spin-up event (or glitch) is believed to result from the hydrodynamic relaxation of the crust and fluid according to the Ekman pumping process analyzed in this paper \cite{van10}.
In Figure \ref{epfig5}, we plot the observed spin-down curve of the Vela neutron star after its 1985 glitch.
The glitch recovery has been fitted to a function of the form \cite{mcc87}
\begin{equation}
 \Omega(t)=c_1 e^{\varpi_+^* t}+c_2 e^{\varpi_-^* t} +c_3 t +c_4  \,, \label{ep501}
\end{equation}
with $c_3=0$ to give $c_1=(1.73\pm0.01)\times10^{-5},\,c_2=(4.1\pm0.6)\times10^{-7},\,c_4=(9.475\pm0.006)\times10^{-5},\,\varpi_+^*=(2.19\pm0.05)\times10^{-7},\,\varpi_-^*=(1.1\pm0.1)\times10^{-5},\,\Omega_s=70.4$ (all in units ${\rm rad\,s^{-1}} $),
and $\left( d \Omega/ d t\right)_{t=0}=-8.69 \times 10^{-7}\,{\rm rad\,s^{-2}}$.
In a neutron star, $K$ and $\rho_n$ are not known.
The only restriction is that (\ref{ep319})--(\ref{ep321}) must be positive definite.
Choosing $K=0.001$ and $\rho_n=0.1$, we deduce $I/J=6.898$ and $\beta E^{1/2}=2.75\times10^{-6}$ from (\ref{ep319})--(\ref{ep321}).
Using Figure \ref{epfig3}, we find $B=4.20\times10^{-6}$ and $B'=-1.05$.
For these values, we calculate $J=0.120$, and hence, from (\ref{ep319}), we obtain $E=1.05\times10^{-5}$.
The extracted value of $E$ is roughly consistent with the theoretical expectations outlined in the previous paragraph, however $B$ and $B'$ are not.
The agreement is unsurprising in view of the simplifying assumptions made, e.g., cylindrical geometry and initial co-rotation between the fluid components.
Glitch recovery in neutron stars has been explored more fully in a recent paper, where an analytic solution in a sphere is applied \cite{van10}.
However, because neutron stars are strongly stratified, the extracted parameters must be treated as body averaged quantities, an issue which does not affect terrestrial experiments.

\section{Conclusions}
\label{epsec5}

We present a recipe for interpreting non-invasively the transport processes at work in superfluid spin-up experiments like those pioneered by the Tsakadze group \cite{tsa80}, whose observed features have resisted theoretical explanation until recently.
The interpretative recipe is sensitive to certain physics, e.g., Ekman pumping and superfluid turbulence, that does not occur normally in second-sound or torsional-oscillator experiments, which rely on small amplitude vibrational modes.
By analyzing the response of a superfluid-filled cylinder to an impulsive initial acceleration, the recipe also yields unique values for the transport coefficients $B$, $B'$ and $\eta$ in general, two relations between the three coefficients in the strong coupling regime, and can be extended to vessels of arbitrary shape \cite{van11a,van11b}.
Indeed, in the strong coupling regime, two identical spin-up experiments in differently shaped containers can be enough to lift the degeneracy between $B$, $B'$ and $E$ inherent in a single experiment (see section \S\ref{epsec3c} and \S\ref{epsec4a}).
The method cannot compete with traditional measurements of $B$, $B'$  and $E$ in He II but may help supplement other techniques in $^3$He and $^3$He-$^4$He mixtures, where accuracies at present are lower.
The method is a powerful tool in interpreting the spherical experiments of Tsakadze and Tsakadze \cite{tsa72,tsa73,tsa80}, and glitch recoveries in neutron stars \cite{van10}, where only remote observations are possible.  \newline

CAVE acknowledges the financial support of an Australian Postgraduate Award and the Albert Shimmins write-up award.
CAVE also thanks the anonymous referees for their constructive suggestions on the manuscript.

\appendix
\section*{Appendix}

For a container of arbitrary axisymmetric shape, the fluid is contained within the volume $-h(r)<z<h(r)$, where $(r,z)$ refer to cylindrical coordinates, scaled by the length-scale of the container $L$.
Equation (\ref{ep301}) holds, with the modified definitions
\begin{eqnarray}
 K&=&I_f^*/I_c\,,\label{tsa201}\\
 g^A(\tau)&=& -\frac{4 \pi }{I_f}\int_0^R \frac{{\rm dr}\,r^3 J(r) \left[  e^{\omega_+(r)\tau}-e^{\omega_-(r)\tau} \right]}{\omega_+(r)-\omega_-(r)} \,,  \label{tsa204} \\
 g^B(\tau)&=& -\frac{4 \pi }{I_f}\int_0^R \frac{{\rm dr}\,r^3 \left\{ \omega_-(r) \left[e^{\omega_+(r)\tau}-1\right]-\omega_+(r)\left[ e^{\omega_-(r)\tau}-1\right] \right\} }{\omega_+(r)-\omega_-(r)}  \,, \label{tsa205} \\
 \omega_{\pm}&=&-\frac{1}{2}\left[\beta+\frac{I(r)}{h(r)}\right]\pm \left\{\frac{1}{4}\left[\beta+\frac{I(r)}{h(r)}\right]^2-\frac{\beta  J(r)}{h(r)} \right\}^{1/2} \, ,  \label{tsa206} \\
 \beta&=&\frac{2 B E^{-1/2}}{2-B'} \, , \label{tsa207}\\
 I(r)&=&\frac{1}{\lambda_{-}(r)\left[\lambda_{-}^2(r)+\lambda_{+}^2(r)\right]  } \nonumber \\
 &&\times \left\{\left[\frac{1-H^4(r)}{1+H^4(r)}\right]^2\left[\lambda_{+}^2(r)-\lambda_{-}^2(r)\right]^2+4\lambda_{+}^2(r)\lambda_{-}^2(r)\right\}^{1/2}\, , \label{tsa208}\\
 J(r)&=&\frac{\rho_n H^2(r)}{2 \rho \lambda_{-}(r) \left[1+H^4(r)\right]} \nonumber \\
 &&\times \left\{ \left[\lambda_{-}^2(r)+\lambda_{+}^2(r)\right]\left[1-H^4(r)\right]+4 \lambda_{-}^2(r) H^4(r)\right\} \, , \label{tsa209} \\
 \lambda_{\pm}(r)&=&\frac{\rho^{1/2}}{H(r)}\left\{ \left[\frac{\left(B'-2\right)^2+B^2}{\left(\rho_n B'-2 \rho \right)^2+\left(\rho_n B\right)^2}\right]^{1/2} \right. \nonumber \\
 && \left.  \mp \frac{2 \rho_s   B \left[1+H^4(r)\right] }{H^2(r)\left[\left(\rho_n B'-2\rho\right)^2+\left(\rho_n B\right)^2\right] }\right\}^{1/2} \,, \label{tsa210} \\
 H(r)&=&\left[1+\left(\frac{d h(r)}{dr}\right)^2\right]^{1/4}\,,\\
 I_f&=&\frac{I_f^*}{\rho L^5}=4\pi\int_0^R {\rm dr} r^3 h(r)\,,
\end{eqnarray}
where $I_f^*$ is the scaled moment of inertia of the contained fluid (as if it were rotating as a rigid body) and $R$ is the cylindrical radius of the vessel, scaled to $L$.
In a cylinder, one has $h(r)=1$, $L=h$, $R=r/h$, and we recover (\ref{ep301a})--(\ref{ep308}).

\bibliographystyle{spphys}       % APS-like style for physics
\bibliography{extparam}   % name your BibTeX data base

\end{document}